\documentclass[%
 reprint,
superscriptaddress,
 amsmath,amssymb,
 aps,
pra,
]{revtex4-2}
\usepackage{float}
\usepackage{subfigure}
\usepackage{xcolor}
\usepackage{graphicx}
\usepackage{dcolumn}
\usepackage{bm}
\usepackage{physics}
\usepackage{siunitx}
\usepackage{ulem}

\begin{document}

\preprint{APS/123-QED}

\title{Hiding images in quantum correlations}

\author{Chloé Vernière} \email[Corresponding author: ]{chloe.verniere@insp.jussieu.fr}
 \affiliation{Sorbonne Université, CNRS, Institut des NanoSciences de Paris, INSP, F-75005 Paris, France\
 }
\author{Hugo Defienne}
 \affiliation{Sorbonne Université, CNRS, Institut des NanoSciences de Paris, INSP, F-75005 Paris, France\
 }
\date{\today}

\begin{abstract}

Photon-pair correlations in spontaneous parametric down conversion are ubiquitous in quantum photonics. The ability to engineer their properties for optimising a specific task is essential, but often challenging in practice. We demonstrate the shaping of spatial correlations between entangled photons in the form of arbitrary amplitude and phase objects. By doing this, we encode image information within the pair correlations, making it undetectable by conventional intensity measurements. It enables the transmission of complex, high-dimensional information using quantum correlations of photons, which can be useful for developing quantum communication and imaging protocols.

\end{abstract}

\maketitle

Spontaneous parametric downconversion (SPDC) is the source for many quantum optics experiments. In such a non-linear process, a photon from the pump beam spontaneously splits into two entangled photons of lower frequencies. At the outputs, properties of down-converted photons are set by the type and geometry of the nonlinear crystal, and the pump beam characteristics~\cite{hong1985theory}. The ability to engineer these properties is key for quantum photonics applications.

Many techniques have been developed to control correlations and entanglement of down-converted photons. Engineering the crystal parameters, such as its poling profile and overall geometry, allows for example to control spectral and spatial properties of the pairs~\cite{branczyk2011engineered,dosseva2016shaping,yesharim2023direct}. Another well-established way to manipulate their properties is to shape the pump beam. Although a few works report on spectral correlation control~\cite{valencia2007shaping,hendrych2007tunable}, the majority focus on manipulating spatial correlations between photon pairs by modulating the pump's spatial profile. This is typically done using spatial light modulators (SLMs). Experimental demonstrations involve structuring the pump in the form of discrete spatial mode carrying orbital angular momentum (OAM) to influence correlations in the corresponding bases~\cite{mair2001entanglement,walborn2003multimode,kovlakov2017spatial,kovlakov2018quantum,d2021full,jabir2017direct,bornman2021optimal}, modulating its spatial phase in the crystal plane to engineer momentum correlations~\cite{minozzi2013optimization,Boucher_pumpshaping,lib2020spatially}, controlling its spatial coherence to tune the degree of entanglement~\cite{defienne2019spatially,zhang2019influence} and even actively compensating for scattering~\cite{lib2020real}. 

Using SLMs, however, the complexity of the correlation structures that can be generated is limited. Interestingly, two studies conducted in different contexts - one to study coherence transfer~\cite{monken_1998} and the other to investigate spatial resolution enhancement~\cite{Unternahrer:18} - have indirectly modulated correlations between pairs without using an SLM but placing simple amplitude objects in the pump path. In our work, we generalize this concept and demonstrate that momentum correlations can be shaped under the form of arbitrary amplitude and phase objects. By doing so, images are encoded in the correlations between photons and are not detectable by conventional intensity measurements.   

\begin{figure}
    \centering
    \includegraphics[width=1\columnwidth]{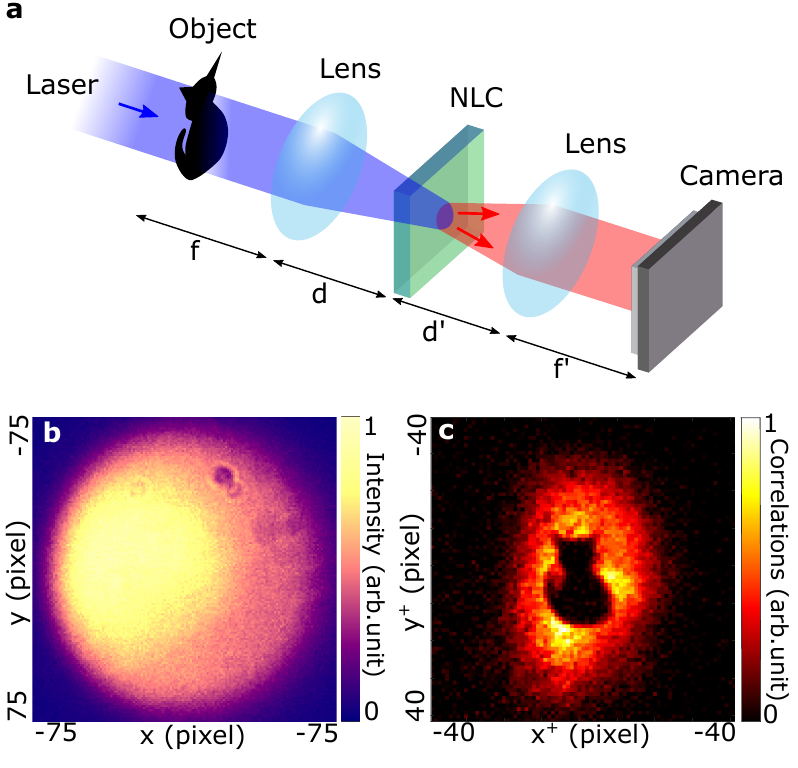}  
    \caption{\textbf{Encoding images in photons correlations.} \textbf{a,} Simplified experimental scheme. A continuous wave laser at $405$ nm illuminates an object (standing cat-shaped mask). It is Fourier imaged onto a thin non-linear crystal (NLC) by a lens $f$. Spatially-entangled photon pairs produced by spontaneous parametric down-conversion (SPDC) around $810$nm are detected in a Fourier plane of the crystal using a lens ($f'$) and a single-photon sensitive camera. The distance between the object and the first lens, and this between the second lens and the camera must be set to $f$ and $f'$, respectively. $d \neq f$ and $d \neq f'$ in general.  \textbf{b,} Conventional output intensity image $I(x,y)$. \textbf{c,} Sum-coordinate projection $\Gamma^+(x^+,y^+)$ of the second-order spatial correlation function $G^{(2)}$, referred to as the 'correlation image'. $\num{9e6}$ frames were acquired using an Electron Multiplied Charge Couple Device (EMCCD) camera and with the specific lenses arrangements detailed in Figure~\ref{fig:set-up}.a. and Figure S1.}
    \label{Figure1}
\end{figure}

Our approach is based on a key property of SPDC, which involves the complete transfer of the pump's spatial coherence to that of the down-converted field~\cite{jha2010spatial}. As shown in Figure~\ref{Figure1}.a, photon correlations are shaped by projecting the Fourier spectrum of a given object $t$ onto a thin non-linear crystal. The down-converted field produced in the crystal plane has a transverse spatial shape identical to that of the object's Fourier transform. After the crystal, it propagates as a spatially incoherent field, leading to an intensity $I$ that remains flat and constant in the far-field (Fig.~\ref{Figure1}.b). By looking at the second-order coherence, however, the correlation function is 
\begin{equation}
\small G^{(2)}(\vb{r_-},\vb{r_+})= \left| \text{t}\left(-\frac{f}{f'} \vb{r_+}  \right)\right|^2,
\label{eq:G2}
\end{equation}
where $f$ (resp. $f'$) are the focal lengths of the lenses before (resp. after) the crystal, $\vb{r_-} = \frac{1}{2}({x_1-x_2},{y_1-y_2})$ and $\vb{r_+} = \frac{1}{2}({x_1+x_2},{y_1+y_2})$ are defined as the minus and sum-coordinates, respectively, and $x_1$, $y_1$, $x_2$ and $y_2$ are the transverse spatial coordinate associated with each photon. To reveal the object, one can measure $G^{(2)}$ and sum it along the minus-coordinate axis i.e.  $\Gamma^+(\vb{r_+}) = \int G^{(2)}(\vb{r_-},\vb{r_+}) d\vb{r_-} \propto | \text{t} (\vb{r_+}) | ^2 $. For clarity, $\Gamma^+$ will be referred to as the `correlation image' throughout the manuscript. Figure~\ref{Figure1}.c shows such a correlation image revealing the standing-cat shaped object. 
In the following, we provide a comprehensive description of the experimental system used to capture these images. Additionally, we conduct an in-depth analysis of our approach, assessing its compatibility with both phase and amplitude objects, the flexibility of its setup and the quality of the resulting correlation image.

\begin{figure*}
    \centering
    \includegraphics[width=1\textwidth]{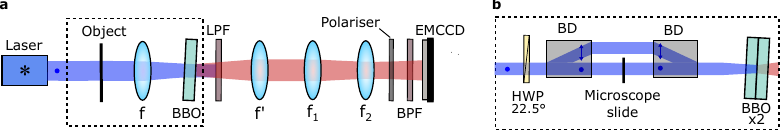}  
    \caption{\textbf{Experimental setup}. \textbf{a,} For imaging amplitude objects, a continuous wave laser emits a collimated beam of horizontally polarized light at $\lambda_p= 405$nm. The beam with a diameter of approximately $2$mm illuminates an amplitude object. The object is positioned in the focal plane of lens $f$ which focuses the beam into a 0.5 mm thick non linear crystal of $\beta$-barium borate (BBO). Within this crystal, photons at $405$nm undergo type-I SPDC, transforming into degenerate spatially entangled photons at $810$mm. A long-pass filter (LPF) filters out the pump beam after the crystal. The generated photons are collected by the lens $f'=35$mm and pass through lenses $f_1$ and $f_2$. An EMCCD camera positioned in the focal plane of $f_2$ detects the photons. A bandpass filter (BPF) at $810 \pm 5$nm and a vertically positioned polarizer eliminate pump residues and ambient light. The distance between each consecutive lens is equal to the sum of each lens's focal length. \textbf{b,} For phase imaging, all the elements located in the black dashed line box are replaced by a interferometric system. A half wave plate (HWP) at $22.5^{\circ}$ is positioned after the laser. Two beam displacers (BD) in calcite are placed in succession to create the two arms of the interferometer, in one of which a microscope slide is inserted to shift the phase. The polariser before the camera is set to $45^{\circ}$. A spatial light modulator (SLM) is inserted in the focal plane between lenses $f_1$ and $f_2$ (not represented). {See also sections I and IV of the supplementary document for more details on the experimental setups.} }
    \label{fig:set-up}
\end{figure*}

A more detailed experimental setup is shown in Figure~\ref{fig:set-up}.a. An amplitude object is illuminated by a blue collimated laser beam at $405$nm. The object is then Fourier-imaged onto a $1$mm-thick $\beta$-barium borate (BBO) crystal. Degenerate spatially entangled photon pairs at $810$nm are produced via type-I SPDC. At the output, the crystal plane is Fourier-imaged onto an Electron Multiplied Charge Coupled Device (EMCCD) camera by $f'$ and further magnified by  $f_1$-$f_2$, allowing photons momenta to be matched to pixel positions. The EMCCD camera is used to perform both intensity and correlations measurement between the incident photons, enabling the computation of $I$ and $G^{(2)}$ for the photons pairs using the method described in Ref.~\cite{defienne2018general}.

\begin{figure}[t]
    \centering
    \includegraphics[width= 1 \columnwidth]{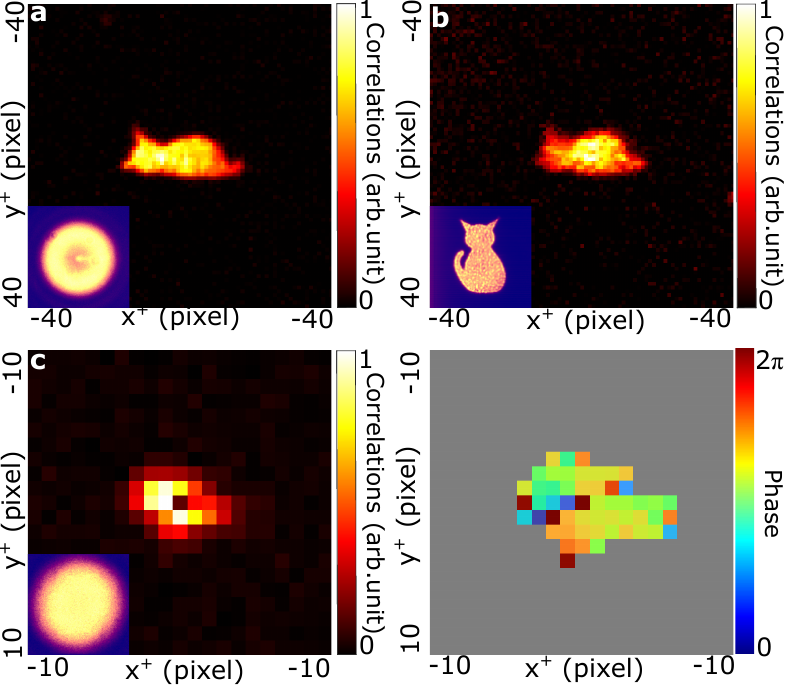} 
    \caption{\textbf{Results with amplitude and phase objects.} \textbf{a, b}, Correlation images obtained using a sleeping cat-shaped mask as the object. The inset at the bottom left shows the near-field intensity image. $\num{3.8e6}$ and $\num{6.3e6}$ frames were acquired.\textbf{(b)}, We also insert a second object with a standing cat shape in the photon pair beam path. This second object was inserted in the Fourier plane between lenses $f_1$ and $f_2$ shown in Figure~\ref{fig:set-up}.a. Correlation image \textbf{(c)} revealing the amplitude shape of a virtual point-like object. Phase of the object \textbf{(d)} reconstructed by phase-shifting holography. The gray area corresponds to undefined phases. $10^7$ frames were acquired. }
    \label{fig:results_tot}
\end{figure}

Figures~\ref{Figure1}.b and c show intensity and correlation images, respectively, obtained using an opaque mask of a standing cat as the object. Similar results are also obtained using a transparent mask with a sleeping cat shape, as shown in Figure~\ref{fig:results_tot}a.  In both cases, the objects are not visible in the direct intensity images ; they can only be unveiled through correlation measurements. This `hiding' of images within photon correlations becomes even more noticeable when the intensity of the photon pair beam is also spatially modulated.
For example, one can insert a second object, such as a mask shaped like a standing cat, positioned at the center of the photon pair beam path, within a Fourier plane between $f'$ and $f_1$. In this case, an intensity measurement reveals an image of the standing cat mask (Fig.~\ref{fig:results_tot}.b, inset), while a correlation measurement reveals that of the sleeping cat (Fig.~\ref{fig:results_tot}.b). 

Our approach also operate with phase objects. To demonstrate this, we use a point-like object i.e. $t(\vb{r}) \approx \delta(\vb{r})e^{i \theta_t(\vb{r})}$, where $\theta_t$ is its phase and $\delta$ the Dirac delta function. It is visible in the correlation image  shown in Figure~\ref{fig:results_tot}.c. In practice, this object is virtual. It is created by removing the lens $f$ from the experimental setup in Figure~\ref{fig:set-up}.a. Indeed, by doing so, the laser illuminating the crystal becomes collimated, equivalent to imaging a point-like object with a width inversely proportional to the beam diameter. Then, a glass slide (microscope slide) is inserted into the pump beam path to set its phase $\theta_t$ to a random arbitrary value.

To retrieve the phase, a common-path interferometer is constructed. Modifications to the experimental setup are shown in Figure~\ref{fig:set-up}b. The collimated pump beam is divided into two optical paths using a half wave plate (HWP) at $22.5^{\circ}$ and two beam displacers (BD). The microscope slide is placed in one of the paths (the imaging path) while the other is used as a reference. A second type-I BBO crystal is added to the setup. It is optically contacted to the first one and rotated $90^{\circ}$. This second crystal interacts with the reference pump to produce photon pairs that are cross-polarized to those generated by the first crystal. Both down-converted fields are Fourier-imaged and superimposed onto the camera. To make them interfere, the polariser is rotated to $45^\circ$. In addition, a spatial light modulator (SLM) is placed between lenses $f_1$ and $f_2$ (not represented). It allows controlling the phase $\theta$ of the vertically-polarized photon pairs, i.e., those generated by the reference pump.

In the camera plane, the second-order correlation function can thus be written
\begin{equation}
\small G^{(2)}(\vb{r_-},\vb{r_+})= \left| t (\vb{r_+}) + e^{i 2 \theta}\right|^2 \propto |\text{cos}(\theta_t ( \vb{r_+})-2 \theta )|^2,
\label{eq:G2}
\end{equation}
where the magnification factor has been omitted for clarity. $\theta_t$ is reconstructed by acquiring $G^{(2)}$ for  four different phase shifts $\theta \in [0, \pi/4,\pi/2,3 \pi /4]$ using the phase-shifting holography formula~\cite{Yamaguchi:97}. Figure~\ref{fig:results_tot}d shows 
an image of the object's phase. It is uniform and non-zero in the region where the object's amplitude is non-zero. Section IV of the supplementary document details the phase-shifting process.

Finally, we analyze some practical aspects of our approach, including the flexibility of its optical arrangement and the trade off between spatial resolution and signal-to-noise (SNR) ratio. As pointed out in Figure~\ref{Figure1}.a, the distance $d$ between lens $f$ and the crystal, and the distance $d'$ between the crystal and lens $f'$, are not constraints. Indeed, it can be shown theoretically that when $d$ and $d'$ differ from $f$ and $f'$, respectively, it only produces an extra phase term on the down-converted field in the camera plane, that is not detected by correlation measurement (see section V of the supplementary document). Experimentally, Figures~\ref{Figure4}.a-c show three correlation images acquired using the setup in Figure~\ref{fig:set-up}.a, with the same object, but for different values of $d$ and $d'$. We observe nearly identical images, showing that, in practice, the distances between the two lenses and the crystal are irrelevant. This is further confirmed by the profiles in Figure~\ref{Figure4}.d. 

Furthermore, Figure~\ref{Figure4}.e shows the variations in spatial resolution and SNR in the correlation image as a function of the focal length $f$. In this context, {we refer to the spatial resolution as} pixel resolution i.e. the total number of pixels over which the image is discretized. It corresponds to the width of the illumination area measured in a correlation image without any object present. The blue curve (left scale) shows the resolution of the system for different lenses $f$ of focal lengths $50$, $75$, $100$, $150$ and $200$mm. The smaller the focal length $f$ is, the wider is the correlation width, with a $1/f$ relationship~\cite{walborn2010spatial}. However, as shown by the black curve (right scale), decreasing $f$ also deteriorates the SNR in the retrieved image (for a fixed acquisition time), which evolves as $f^2$. There is thus a trade-off between the image pixel resolution and its SNR. 

\begin{figure}
    \centering
    \includegraphics[width=1 \columnwidth ]{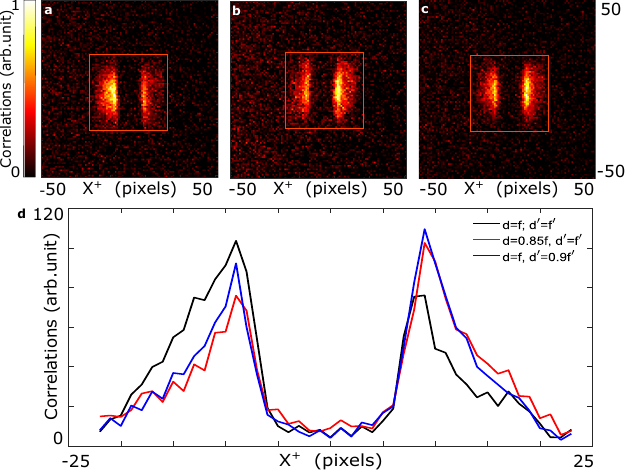}
    \caption{\textbf{Distances independence}. Correlation measurements acquired while varying the distances $d$ and $d'$ such that \textbf{(a)} $d=f$, $d'=f'$; \textbf{(b)} $d=0.85f$, $d'=f'$; \textbf{(c)} $d=f$, $d'=0.9f'$. The object imaged is a resolution grating ($1.25$lp/mm). \textbf{d}, Correlation intensity profile vertically averaged in the region inside the red square in \textbf{a-c}. $\num{6.6e5}$ frames were acquired for each correlation images.}
    \label{Figure4}
\end{figure}

\begin{figure}
    \centering
    \includegraphics[width=1 \columnwidth ]{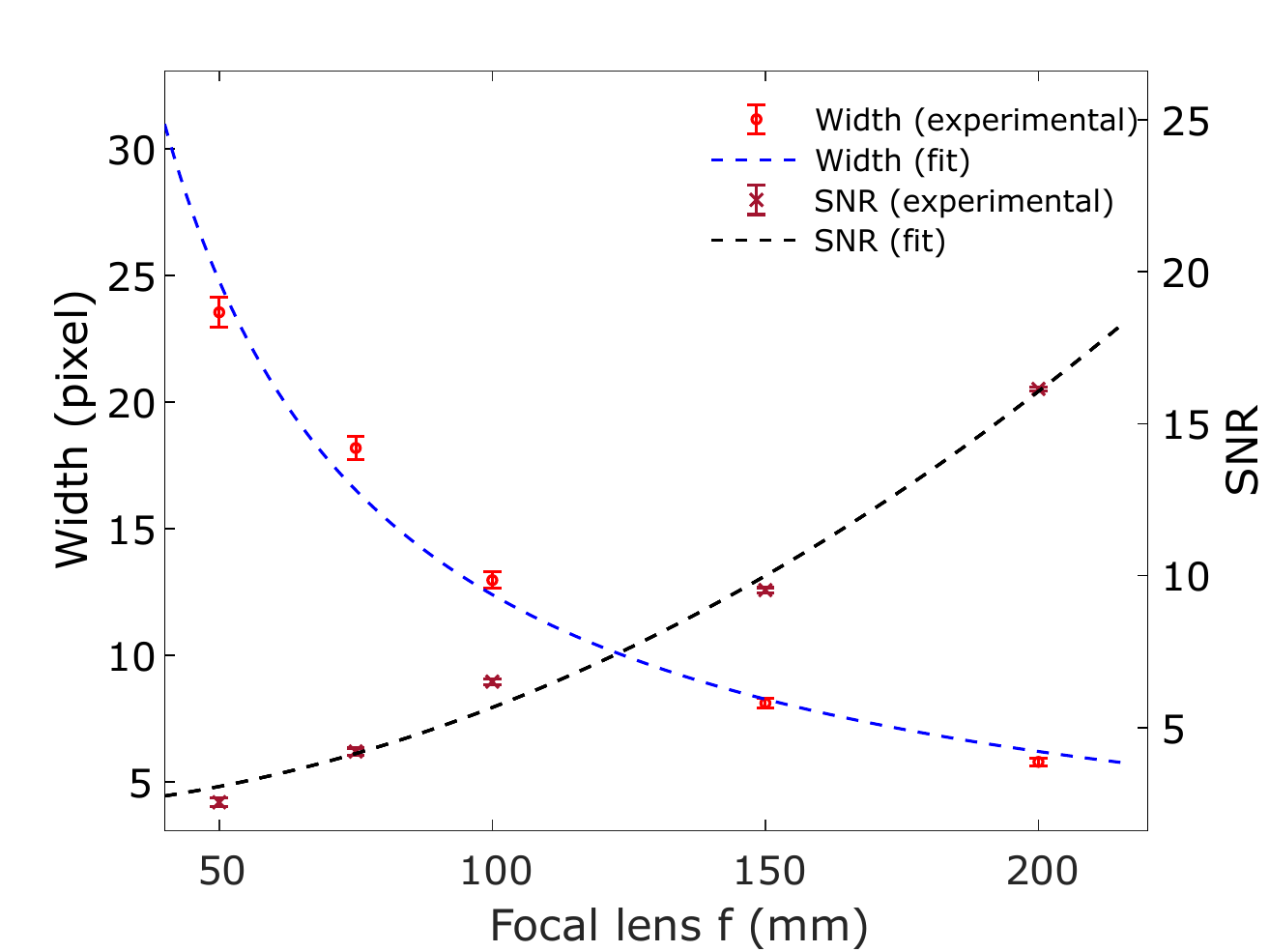}
    \caption{\textbf{Signal-to-noise ratio (SNR) and resolution trade-off}. Left scale: Correlation width as a function of the focal length $f$. Experimental data (red circles) and fit (dashed blue). $\num{6.6e5}$ frames were acquired for all data points, except for $f=50$mm for which $\num{5e6}$ frames were required. Fit according to a $1/f$ relationship as predicted by the theory model. Right scale: SNR (dark red crosses) and fit (dashed black) as a function of the focal length $f$. Fit according to a $f^2$ relationship. $\num{6.6e5}$ frames were acquired for each data points. }
    \label{Figure5}
\end{figure}

\section*{Conclusion}
We have demonstrated the shaping of quantum correlations between photon pairs into amplitude and phase objects. No direct intensity measurement reveals information about the encoded object; only a coincidence measurement does. In addition, the system proves to be flexible in terms of space requirements, as some distances can be compressed, and also in terms of resolution, as the latter can be fully monitored only by $f$. However, one drawback is that the EMCCD camera used here require several hours to acquire the $10^6-10^7$ frames needed to reconstruct the correlation images. To address the issue, a potential solution is using a different single-photon camera technology, such as a time-stamping camera~\cite{Nomerotski_2019,vidyapin2023characterisation}, which combines high resolution and speed.

Capitalizing on the flexibility and experimental simplicity of the setup, this could subsequently enable the development of a new  imaging and encoding-decoding protocols for high dimensional quantum information processing. This could find some applications in quantum communication and cryptography. 

\section*{Acknowledgements}

H.D. acknowledges funding from the ERC Starting Grant (No. SQIMIC-101039375).

\section*{Author Contributions}
C.V. analysed the data, designed and performed the experiments. C.V and H.D. conceived the original ideal, discussed the results and contributed to the manuscript. H.D supervised the project.

\bibliography{bib}

\end{document}


\preprint{APS/123-QED}

\title{Supplementary document: Hiding images in quantum correlations}

\author{Chloé Vernière} \email[Corresponding author: ]{chloe.verniere@insp.jussieu.fr}
 \affiliation{Sorbonne Université, CNRS, Institut des NanoSciences de Paris, INSP, F-75005 Paris, France\
 }
\author{Hugo Defienne}%
 \affiliation{Sorbonne Université, CNRS, Institut des NanoSciences de Paris, INSP, F-75005 Paris, France\
 }
\date{\today}

\maketitle

\section{Details on the experimental setups}

The laser is a 100mW-power continuous-wave vertically polarised diode laser at $402$nm (Coherent). The polarisation can be tuned with a half-wave plate placed in front of the laser. The EMCCD (iXon Life 897, Andor) has pixels of size $16$x$16\mu$m$^2$ and a quantum efficiency of $\sim70\%$ at $810$nm. Its temperature is set to -$60^\circ$C and is operated with a horizontal pixel shift readout rate of $17$ MHz, a vertical pixel shift of $0.3$µs, a vertical clock amplitude voltage of $4$ V above the factory setting, and an amplification gain of $1000$. The exposure time is set to $3$ms. In most of our results, $10^6$-$10^7$ frames were acquired.  

The cat-shaped amplitude objects are imprinted in black on a plastic sheet. Their size is $\sim 3\cross 1.5$mm$^2$. The grating used as an amplitude object is a positive variable line grating test target (Thorlabs, R1L3S6P).

The non-linear crystal is a $0.5\times5\times5$mm type-I $\beta$-barium borate (BBO) crystal (Newlight Photonics). It is slightly tilted around the horizontal axis to ensure near-collinear phase matching of photons at the output. For the phase object, the crystal is replaced by two identical type-I BBO crystals with perpendicular axis to each other and optically contacted. The photon pairs produced are thus entangled both in space and polarisation. The crystals are slightly tilted around the vertical and horizontal axis to ensure near-collinear phase-matching of the photons at the output.

\begin{figure}
    \centering
    \includegraphics[width=\textwidth]{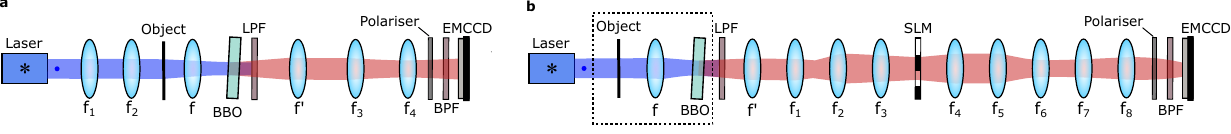}
    \caption{\textbf{Experimental setups}. Two configurations of the setup used for the amplitude objects data acquisition. See Figure \ref{fig:setup_phase}.a for more details for the phase object configuration inside the rectangular dashed box. BBO : non-linear crystal in $\beta$-barium borate. LPF : long-pass filter. BPF : band-pass filter. SLM : spatial light modulator. EMCCD : electron-multiplying charge-coupled device.}
    \label{fig:setups_details}
\end{figure}

Figure~\ref{fig:setups_details} shows the two different setups used in our experiments. Results shown in Figure 1b-c and Figure 3a-b of the manuscript are acquired with the setup presented in Figure~\ref{fig:setups_details}a. Lenses $f_1=150$mm and $f_2=300$mm magnify the laser. $f=100$mm projects the Fourier transform of the object on the crystal. Lenses $f'=35$mm, $f_3=150$mm and $f_4=50$mm do far-field imaging of the crystal on the camera. The effective focal length is $f'_{eff}=\frac{f_4 f'}{f_3}=11.7$mm. All the other results were acquired with the setup shwon in Figure~\ref{fig:setups_details}b. In this configuration, there is no magnification of the laser and $f=100$mm. The photon pairs produced by the crystal are imaged onto an SLM by two telescopes with $f'=35$mm - $f_1=60$mm and $f_2=50$mm - $f_3=125$mm (magnification of $4.3$). They are then de-magnified by two telescopes $f_4=200$mm - $f_5=75$mm and $f_6=125$mm - $f_7=150$mm (magnification of 0.45) and Fourier imaged on the camera with $f_8=50$mm. The effective focal length between the crystal and the camera is $f'_{eff}= \frac{f_8}{0.45.4.3}=26$mm. Unless specified, all lenses are disposed in confocal configuration. The SLM is a phase only modulator (Holoeye Pluto-2-NIR-015) with $1920 \cross 1080$ pixels and
a $8\mu$m pixel pitch. It is here represented in transmission but actually works in reflection. Both setups have a short-pass filter just before the crystal to remove stray light emitted by the laser cavity (non represented), a long-pass filter right after the crystal to filter out the pump beam and a bandpass filter (810nm-10) in front of the camera. 

\section{Theoretical demonstration of $G^2$}

Following Ref.~\cite{Abouraddy:02}, we derive Equation $1$ in a 1D system and then extend the results to a 2D system. Here we use the generic configuration introduced in Figure $1$ of the manuscript, identifying only two lenses in the system, namely $f$ between the object and the crystal, and $f'$ between the crystal and the camera. The crystal is thus illuminated by the Fourier transform of an object described by $t(x)$, such that the illumination of the crystal is 
\[\mathcal{FT}\left[t\right](x) = \int dx'\; t(x') \exp(-2i\pi \frac{x x'}{\lambda_p f})\]
with $\lambda_p=405$nm the pump wavelength.
Both signal and idler have the same impulse-response functions $h(x,x')$, that of a 2-f Fourier system of effective focal length $f'$, which reads  
\[h(x,x') = \exp\left(-i \frac{2\pi}{\lambda f'} x x'\right)\] with $\lambda=2\lambda_p =810$nm the wavelength of the SPDC light. 
The correlation function of the two photons detected by the camera is then given by
\begin{align}
G^{(2)}(x_1,x_2) &= \left| \int dx \; \mathcal{FT}\left[t\right](x) h(x_1,x)h(x_2,x)\right|^2
\nonumber \\
\tiny & = \left| \int dx\; dx'\; t(x')\exp(-2i\pi\left(\frac{x'}{\lambda_p f}+\frac{(x_1+x_2)}{\lambda f'}\right)x) \right|^2
 \nonumber \\
&= \left| \int dx'\;t(x') \delta \left(\frac{x'}{\lambda_p f} + \frac{x_1 + x_2}{\lambda f'}\right)\right|^2
\nonumber \\
&= \left| t(-\frac{f}{f'}x_+)\right|^2.
\end{align}

We note $x_+ = \frac{\lambda_p}{\lambda}(x_1+x_2)= \frac{x_1+x_2}{2}$. More generally, considering an object in $2$D described by the vector $\vb{r} = (x,y)$, we get
\[G^{(2)}(\vb{r}_1, \vb{r}_2) = \left| t\left(-\frac{f}{f'}\vb{r}_+\right)\right|^2.\]

\section{$G^2$ measurement with an EMCCD camera}

As detailed in Ref.~\cite{defienne2018general}, given a set of M frames acquired by the camera, we compute (1) the intensity distribution $I(i,j)$ at pixel $(i,j)$ and (2) the intensity correlations distribution $G^{(2)}(i,j,k,l)$ between pixels $(i,j)$ and $(k,l)$:
\begin{enumerate} 
    \item The intensity distribution is computed by averaging the intensity over all the frames : $I(i,j) = \frac{1}{M}\sum_{m=1}^M I_m(i,j)$, where $I_m(i,j)$ is the intensity value measured at pixel $(i,j)$ of the $m^{th}$ frame.
    \item The second-order intensity correlation distribution between pixels $(i,j)$ and $(k,l)$ is computed by multiplying the values from the same frame of the two pixels and subtracting the product of the value of the two pixels from consecutive frames. This is then averaged over all the frames and reads  
\begin{equation}
\small G^{(2)}(i,j,k,l) = \frac{1}{M}\sum_{m=1}^M I_m(i,j)I_m(k,l) - \frac{1}{M-1}\sum_{m=1}^{M-1} I_m(i,j)I_{m+1}(k,l).
\end{equation} 
The second term enables to subtract the accidental coincidences, \textit{i.e.} photons from two different pairs or from noise. This gives a discrete version of the continuous second-order intensity correlation function $G^{(2)}(\vb{r_1},\vb{r_2})$, where $\vb{r_1}$ and $\vb{r_2}$ are associated to pixel $(i,j)$ and $(k,l)$, respectively.
\end{enumerate}

The sum coordinate projection $\Gamma^+(\vb{x^+},\vb{y^+})$, to which we refer as the 'correlation image', can be computed from $G^{(2)}(\vb{r_1},\vb{r_2})$ by integrating along the minus-coordinate. With continuous variables, 
\begin{equation}
    \Gamma^+(\vb{r_+}) = \int G^{(2)}(\vb{r_-},\vb{r_+})d\vb{r_-}.
\end{equation}
With discrete variables, we in practice compute :
\begin{equation}
    \Gamma^+(k,l) = \sum_{i=1}^{N_x} \sum_{j=1}^{N_y} G^{(2)}(k-i,l-j,i,j).
\end{equation}
$N_x$ (resp. $N_y$) refers to the number of pixels along the horizontal (resp. vertical) direction of the camera sensor.

\section{Measurement with phase objects}

\begin{figure}
    \centering
    \includegraphics[width=0.99\textwidth]{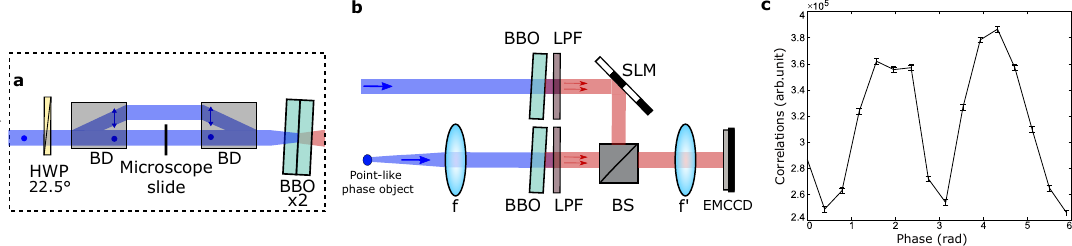}
    \caption{\textbf{Setup details for phase objects.}\textbf{a}, For the phase experiment, elements in the dashed box in Figure \ref{fig:setups_details}.b are replaced by the interferometric system here represented. \textbf{b}, Conceptual scheme of the interferometer. A point-like object is Fourier-imaged onto one crystal that generates photon pairs. This photon pair flux interferes with photon pairs produced by another crystal (used as a reference arm) through a beam splitter (BS). A lens $f'$ images the point-like object onto the camera, and a Spatial Light Modulator (SLM) positioned on the reference arm enables phase-shifting holography. \textbf{c}, Values of $\Gamma^+(\vb{r_+}=0$) in function of the phase $\theta$ programmed on the SLM. $\num{4.2e5}$ frames were taken for each data point.  HWP : half-wave plate. BD : beam displacer in calcite. BBO : non-linear crystal in $\beta$-barium borate. PBS : polarized beam splitter. LPF : long-pass filter. BS : beam splitter. SLM : spatial light modulator.}
    \label{fig:setup_phase}
\end{figure}

To retrieve the phase of an object, we modify the set-up as illustrated in Figure~\ref{fig:setup_phase}.a. A half-wave plate at $22.5^\circ$ sets the pump polarisation to be diagonally polarized and two beam displacers in calcite are inserted. They separate the horizontal and vertical polarisation components of the pump, thus effectively creating the two arms of an interferometer. In one of the arms, a microscope slide is introduced to shift the phase. The two arms then recombine and the beam is incident on the non-linear crystals. The two crystals have their axes perpendicular to each other. The pairs are produced by type-I SPDC and are thus cross-polarized at the output (vertical and horizontal). Only the pairs with horizontal polarization carry the phase shift information. The photon pairs then interfere before the camera using a polarizer rotated to $45^\circ$. This configuration is a common-path interferometer i.e. pairs of photons emitted by each crystal propagate along the same optical path and they interfere only just before the camera. 

To clarify its operation, our experiment can be conceptually represented by a more general interferometric diagram in which there are two separate crystals and their pairs are recombined by a beam-splitter, as illustrated in Figure~\ref{fig:setup_phase}.b. The situation unfolds precisely as if we were employing the setup illustrated in Figure 1 of the manuscript to image a point-like object and using an external reference arm to measure its phase. By inserting or not the microscope slide, the phase of such a virtual point-like object can be changed.

To measure the phase of such a point-like object, we use a phase-shifting holography method. To achieve this, it is necessary to precisely control the phase of the reference arm, i.e., the phase of vertically polarized photon pairs. For this purpose, a SLM is positioned along the path of the photons between the crystals and the camera, as shown in Figure~\ref{fig:setups_details}.b. The SLM enables the modulation of the phase only for vertically polarized photons. At the camera, the resulting second-order correlation function measured is 
\begin{equation}
\small G^{(2)}(\vb{r_-},\vb{r_+})= \left| t (\vb{r_+}) + e^{i 2 \theta}\right|^2 \propto |\text{cos}(\theta_t ( \vb{r_+})+2\theta- \theta_{ref} )|^2,
\label{eq:G2}
\end{equation}
where $\theta_t$ is the phase of the object, $\theta$ is the phase displayed by the SLM and $\theta_{ref}$ the reference phase. The presence of this phase term is a consequence of the fact that the interferometer is not balanced. As we will see bellow, it will be removed by subtraction. Figure~\ref{fig:setup_phase}.c shows oscillations of $\Gamma^+$ at $\vb{r_+}=\vb{0}$ when scanning the phase from $0$ to $2 \pi$ using the SLM.

\begin{figure}
    \centering
    \includegraphics[width=0.99\textwidth]{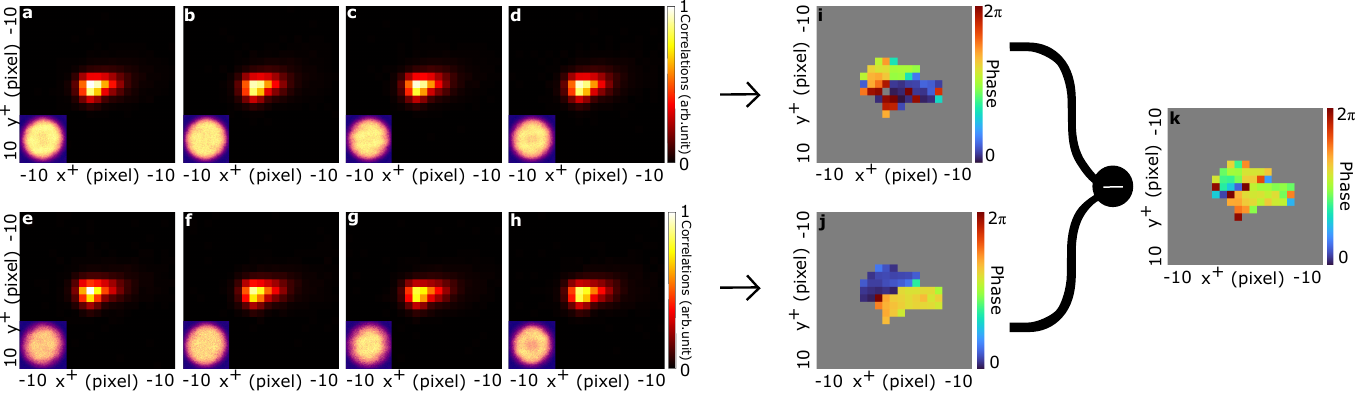}
    \caption{\textbf{Phase computation with four phase holographic measurement.} \textbf{a-d} Correlation images for $\theta \in [0, \pi/4, \pi/2, 3\pi/4]$ without any object in the setup. In inset, corresponding direct intensity image. \textbf{i}, Reconstructed reference phase $\theta_{ref}$ of the correlation image . \textbf{e-h}, Correlation images for $\theta \in [0, \pi/4, \pi/2, 3\pi/4]$ with a phase object in the setup. In inset, corresponding direct intensity image. \textbf{j}, Reconstructed phase $\theta_t-\theta_{ref}$ of the correlation image with the object. \textbf{k}, Reconstructed phase of the object $\theta_t$ by subtracting \textbf{i} to \textbf{j}. The
gray area corresponds to undefined phases.}
    \label{fig:phase_computation}
\end{figure}

In practice, to retrieve the phase of the object, it is sufficient to measure $G^{(2)}$ for only $4$ values of the SLM phase i.e. $\theta \in [0, \pi/4, \pi/2, 3\pi/4]$. This method is called phase-shifting holography~\cite{Yamaguchi:97}. It has been used previously with photon pairs in Refs.~\cite{camphausen2021quantum,Defienne_2022}. By applying phase-shifts $\theta \in [0, \pi/4, \pi/2, 3\pi/4]$ and acquiring the $G^{(2)}$, we retrieve the phase $\theta_t$ by combining the sum-projections as follow :
\begin{equation}
\theta_t - \theta_{ref} = \textrm{arg}(\Gamma^+_{\theta=0}-\Gamma^+_{\theta= \pi/2}+i(\Gamma^+_{\theta=\pi/4}-\Gamma^+_{\theta=3\pi/4})).
\label{eq:holography}
\end{equation}
To eliminate the reference phase $\theta_{ref}$, it is necessary to calibrate the system beforehand by taking a reference measurement without the object, i.e., without the microscope slide. Figure~\ref{fig:phase_computation}.a-d show the four correlation images for $\theta \in [0, \pi/4, \pi/2, 3\pi/4]$ without the microscope slide in the setup used to retrieve the reference phase $ \theta_{ref}$ (Fig.\ref{fig:phase_computation}.i). Figure~\ref{fig:phase_computation}.e-h show the four correlation images with the microscope slide in the setup used to retrieve $\theta_t-\theta_{ref}$ (Fig.\ref{fig:phase_computation}.k). Finally, subtracting the first phase measurement to the second one results to that of the object $\theta_t$, shown in Figure~\ref{fig:phase_computation}.k and Figure $3$d of the manuscript. \num{7.11e6} and \num{15.9e6} frames where acquired for each measure of the holography procedure for respectively the reference and the object measurements, which in total sums to respectively $160$ and $72$ hours of acquisition.

\section{Theoretical demonstration of independance of $G^{(2)}$ with f and f'}

In this section we show that the correlations do not depend on the distances between the crystal and its two neighboring lenses $f$ and $f'$. The derivation is conducted in 1D then extended to the 2D case. We here follow the generic formalism introduced in Figure $1$ of the manuscript. Let $f$ (resp. $f'$) be the distance between the object (resp. the camera) and the lens $L$ (resp. $L'$). Let $d$ (resp. $d'$) be the distance between the crystal and the lens $L$ (resp. $L'$). The input on the crystal then reads 
\[ \mathcal{FT}\left[t\right](x') = \int dx \; t(x) \exp(i\frac{\pi}{\lambda_p}\left(\frac{x^2}{f}+\frac{x'^2}{d}-\frac{(x' f+xd)^2}{f^2d}\right))\]
and the impulse-response function is
\[h(x',x) = \exp(i\frac{\pi}{\lambda}(\frac{x^2}{d'}+\frac{{x'}^2}{f'}-\frac{(x' d'+xf')^2}{d' f'^2})).\]
Then :

\begin{flalign*}
&G^{(2)}(x_1,x_2) = \left|\int dx' \; \mathcal{FT}\left[t\right](x') \; h(x_1,x')h(x_2,x') \right|^2 &\\ 
\end{flalign*}
\begin{multline*}
G^{(2)}(x_1,x_2) = |\iint dx \; dx' \; t(x)\; \exp(i\frac{\pi}{\lambda_p}(\frac{x^2}{f}+\frac{x'^2}
{d}-\frac{(x' f+xd)^2}{f^2d})) \\
\exp(i\frac{\pi}{\lambda}(\frac{x'^2}{d'}+\frac{{x_1}^2}{f'}-\frac{(x_1 d'+x'f')^2}{d' f'^2})) \exp(i\frac{\pi}{\lambda}(\frac{x'^2}{d'}+\frac{{x_2}^2}{f'}-\frac{(x_2 d'+x'f')^2}{d' f'^2})) |^2  
\end{multline*}
\begin{flalign}
&G^{(2)}(x_1,x_2) = \left| \exp(i\frac{\pi}{\lambda_p}(\frac{1}{2f'}-\frac{d'}{2f'^2})(x_1^2+x_2^2)) \int dx\; t(x) \; \exp(i\frac{\pi}{\lambda_p}x^2(\frac{1}{f}-\frac{d}{f^2}))\; \int \; dx'\exp(-i\frac{\pi}{\lambda_p}\frac{2x'}{f}(x+\frac{f}{f'}\frac{x_1+x_2}{2}))  \right|^2 
&\\
&G^{(2)}(x_1,x_2) = \left|\int dx \; t(x)\; \exp(i\frac{\pi}{\lambda_p}x^2(\frac{1}{f}-\frac{d}{f^2}))\delta(x+\frac{f}{f'}\frac{x_1+x_2}{2})\right|^2 
&\\
&G^{(2)}(x_1,x_2) = \left| t(-\frac{f}{f'}\frac{x_1+x_2}{2})\exp(i\frac{\pi}{\lambda_p}(-\frac{f}{f'}\frac{x_1+x_2}{2})^2(\frac{1}{f}-\frac{d}{{f^2}}))\right|^2 
&\\
&G^{(2)}(x_1,x_2) = \left| t(-\frac{f}{f'}x_+)\right|^2.
\end{flalign}

The phase factors associated to $d$ and $d'$ in B7 and B9 do not modify the correlations measured on the camera. In B7, we use $2\lambda_p = \lambda$.

\section{Fitting process in Figure $5$ of the manuscript}
Figure~\ref{fig:correlation_width} shows the correlation images acquired to compute the signal-to-noise ratio (SNR) and the correlation width for different values of the focal length $f$. All were acquired with the same number of frames ($\num{6.6e5}$). The inset in Figure~\ref{fig:correlation_width}.e shows a longer acquisition ($\num{4.9e6}$ frames) to be able to fit the width.

\subsection{Correlation width} 

\begin{figure*}
    \centering
    \includegraphics[width= \textwidth ]{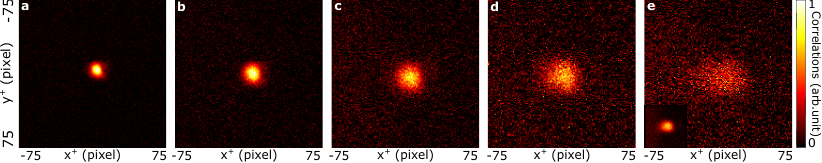}
    \caption{\textbf{Width of the correlations}. Correlations measurements with focal length of the lens before the crystal of (\textbf{a}) $f=200$mm; (\textbf{b}) $f=150$mm; (\textbf{c}) $f=100$mm; (\textbf{d})$f=75$mm; (\textbf{e}) $f=50$mm. All were acquired with $\num{6.6e5}$ frames. \textbf{e} shows in inset a longer acquisition of $\num{4.9e6}$ frames for $f=50$mm to be able to fit the width of the correlations.}
    \label{fig:correlation_width}
\end{figure*}

To compute the width of the correlations, we fit the correlation spots according to a double Gaussian model~\cite{moreau2012realization,edgar2012imaging}. The object $t(\vb{r})$ is in this case the laser itself, which we model as a two dimensional Gaussian \[t(x,y) \propto \exp(-\frac{(x-x_0)^2}{2\sigma_x^2})\exp(-\frac{(y-y_0)^2}{2\sigma_y^2}),\]  where $\sigma_x$ (resp. $\sigma_y$) is the width of the laser along the horizontal (resp. vertical) axis. We subsequently fit the correlation spot according to \[|t(\frac{f}{f'}x_+,\frac{f}{f'}y_+)|^2  \propto \exp(-\frac{(x_+-\Tilde{x}_0)^2}{\Tilde{\sigma}_x^2})\exp(-\frac{(y_+-\Tilde{y}_0)^2}{\Tilde{\sigma}_y^2}),\]where $\Tilde{\sigma}_x = \frac{f'}{f}\sigma_x$ and $\Tilde{\sigma}_y= \frac{f'}{f}\sigma_y$ are the theoretical values for the width of the correlation spot. The values presented in Figure $5$ of the manuscript correspond to the fitted value of $\Tilde{\sigma}_x$ for the data shown in Figure~\ref{fig:correlation_width}a-d,e (inset). Fitting the evolution of the correlation width  as $a/f$, we get $a_{exp} = \num{1.24e3}$ pixel.mm which is close within $5\%$ to the theoretical one $a_{th} = \num{1.30e3}$ pixel.mm. 

\subsection{SNR} 

We define the SNR as the ratio between the mean value of the correlation spot and the standard deviation of the noise around the correlation spot. This standard deviation is proportional to that of the noise of the signal acquired in the same experimental conditions, such that this computation of the SNR actually corresponds to the ratio between the signal and its noise. Experimentally, the correlation spot spreads proportionally to $1/f^2$, and consequently, the signal scales as $f^2$. At a constant acquisition time, and therefore with constant measurement noise, the SNR evolves as $f^2$~\cite{reichert2018optimizing}.

\bibliography{bib}